\documentclass[twocolumn,twoside,slac_two]{revtex4}
\usepackage{graphicx}
\usepackage{fancyhdr}
\begin{document}

\title{Top Physics at the LHC}

%

\author{P. de Jong}
\affiliation{Nikhef, P.O. Box 41882, NL-1009 DB Amsterdam, the Netherlands}

\begin{abstract}
The LHC will be a top quark factory. In this note, the central role
of the top quark for LHC physics will be discussed, and an overview
will be given of the studies of top quark properties in
preparation, with an emphasis on the systematic uncertainties that will
dominate most measurements.
\end{abstract}

\maketitle

\thispagestyle{fancy}


\section{Introduction}

The Large Hadron Collider at CERN is scheduled to deliver its first 
proton-proton collisions in the autumn of 2009. The ATLAS and CMS
experiments are in the final phase of installation and commissioning of
their detectors, and are preparing for measurements of the first collisions.

The LHC will be a top quark factory, both for top-quark pair production
and single top-quark production. In this note, the central role of top
quark production for LHC physics will be explained, and an overview will
be given of the studies of top quark properties in preparation.
Being the only fermion with a Yukawa coupling to the Higgs boson of 
${\cal{O}}(1)$, the top quark plays a central role in almost all models 
of new physics. Reviews of top quark physics at hadron colliders can
be found in Refs.~\cite{quadt,bernreuther,han}.

Lacking data, all results discussed in this note are expectations based
upon Monte Carlo simulations with realistic detector response and 
backgrounds, as documented by CMS~\cite{cmstdr} and ATLAS~\cite{atlascsc}.

\section{Top production}

\begin{figure}[htb]
\centering
\includegraphics[width=80mm]{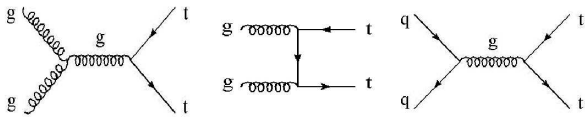}
\caption{Top-pair production diagrams.} 
\label{fig:toppairdiag}
\end{figure}

Figure~\ref{fig:toppairdiag} shows the leading order diagrams for
top-pair production in QCD.
At the LHC, top-quark pair production is dominated by gluon-gluon
fusion (~90\%). The cross-section increase with respect to the Tevatron
is very large, a factor of almost 150. The backgrounds increase as well,
but with a smaller factor (about 10 for W/Z + jets), making the 
signal to background ratio better at the LHC.
On the other hand, there is now
considerable phase space for radiation of extra jets.

In the Standard Model (SM), each top quark decay would produce a W and a
lighter quark (usually a $b$-quark), and the W can decay either leptonically
(to $e$, $\mu$ or $\tau$ plus corresponding neutrino), or to two jets.
This leads to an overall branching fraction of top-quark pairs to
an all-hadronic $bbjjjj$ final state of 46.2\%, to a semi-leptonic
$bb \ell \nu jj$ final state of 43.5\% (29\% for $\ell = e, \mu$ only),
and to a di-leptonic $bb \ell \nu \ell \nu$ final state of 10.3\%
(4.6\% for $\ell = e, \mu$ only).

Recently three new calculations of the top-pair production cross-section
have been published~\cite{moch,cacciari,kidonakis}, each making significant 
steps towards a future full NNLO
calculation. The central values of these calculations at the LHC are in
good agreement, even though there is some difference in the treatment of the 
scale uncertainties. Taking the conservative value of Ref~\cite{cacciari},
the top-quark pair cross-section at $\sqrt{s} = 14$ TeV, using the 
CTEQ6.5 parton distribution functions (pdf), is calculated to be:
$\sigma (t \bar{t}) = 908 \pm 83 \, \mathrm{(scales)} \, \pm 30 \,
\mathrm{(pdf)}$ pb, for a top mass $m_t$ of 171 GeV.
The difference between the predictions using the MRSTW-06 and CTEQ6.5 pdf's 
(53 pb), however, is larger than the pdf uncertainties as evaluated 
within CTEQ (30 pb) and MRSTW (12 pb) individually. 
At $\sqrt{s} = 10$ TeV, the startup centre-of-mass energy of the LHC,
$\sigma (t \bar{t}) \approx 400$ pb.

\begin{figure}[htb]
\centering
\includegraphics[width=80mm]{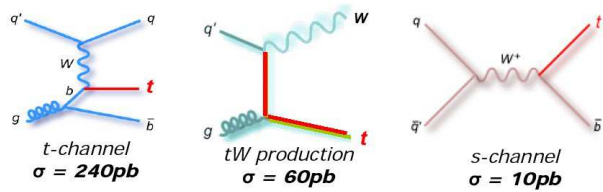}
\caption{Single top production diagrams.} \label{fig:singletopdiag}
\end{figure}

Single top-quark production is an electroweak process, usually divided
into three mechanisms: the t-channel mechanism with the largest expected
cross-section, associated tW production, and the small s-channel (or W*)
production, as shown in Fig.~\ref{fig:singletopdiag}.

\section{Role of the top at the LHC}

At the LHC, the top quark has many hats. It is possible to select
clean samples of top quarks, which can be used to check and calibrate
the detectors, as described in the next subsection.
A precise measurement of the top
production cross-section is a test of advanced QCD resummation
techniques that are also applicable in other calculations. Precise
measurements of top properties can give insight into new physics
beyond the SM affecting top couplings and decay modes. New physics can
also affect top production, for example via top production from decay
of heavy resonances. Finally, SM top quark production is often a major
background in searches for new phenomena like supersymmetry, and it
is important to understand all aspects of top production from data.

\subsection{Top as a tool}

Pure samples of top-pair events can be used to verify trigger
and lepton identification, measure the light and $b$-jet energy scale,
check the missing $E_T$ reconstruction, and measure the $b$-tagging efficiency.
Furthermore, with these events it is possible to study efficiencies
and fake rates in a ``busy'' environment.
The top at the LHC is thus a prime multifunctional tool (yesterday's
sensation, today's calibration!).

\section{Observation and cross-section}

Due to this special role of the top quark in detector commissioning,
CMS and ATLAS aim for a rapid rediscovery of the top quark in early
data. A measurement of the cross-section in $<1$ fb$^{-1}$ would provide 
an interesting early physics result, and is important for searches
for new physics.

\subsection{$t \bar{t}$ production}

\begin{figure}[tbh]
\centering
\includegraphics[width=38mm,height=35mm]{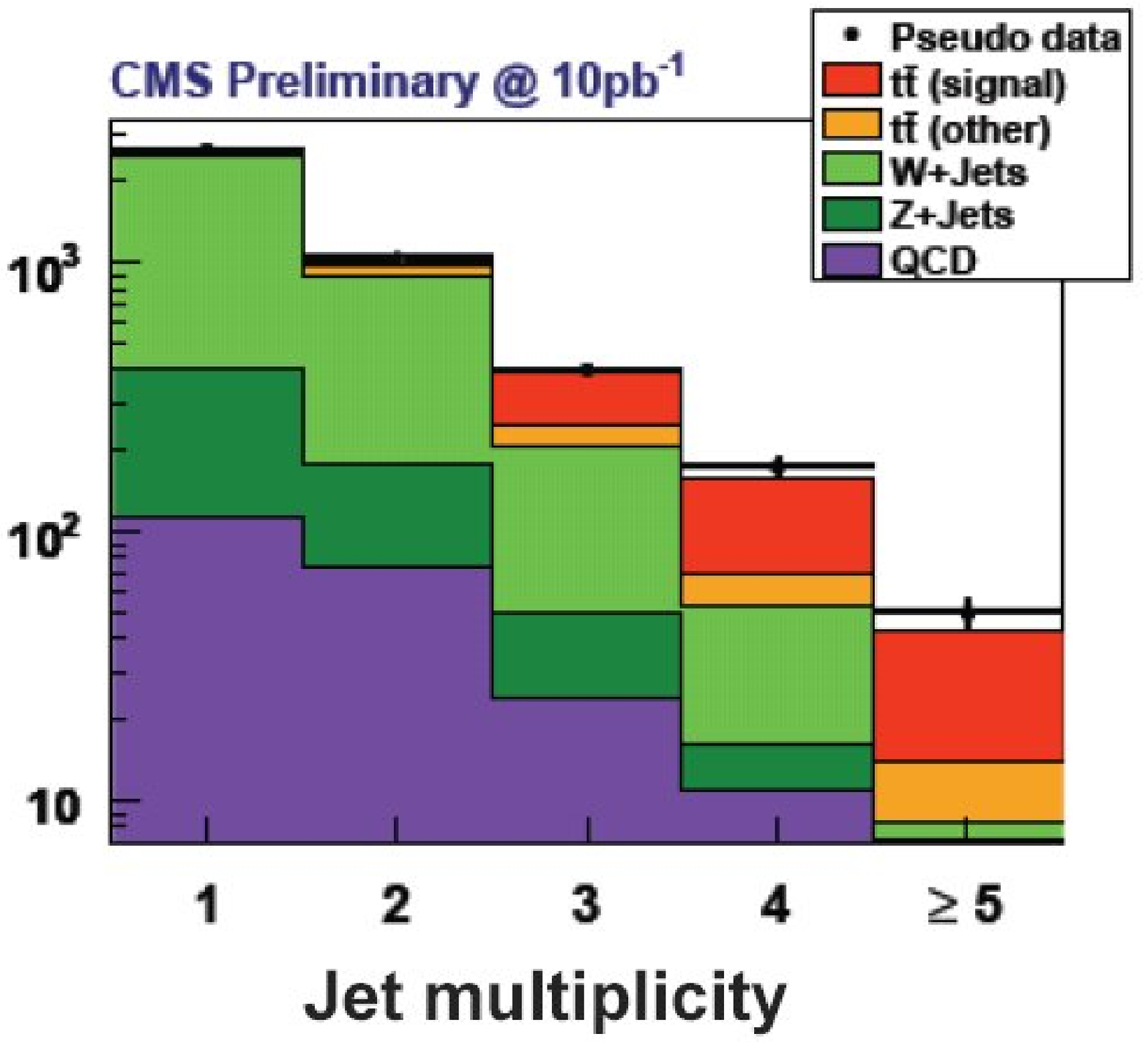}
\includegraphics[width=38mm,height=35mm]{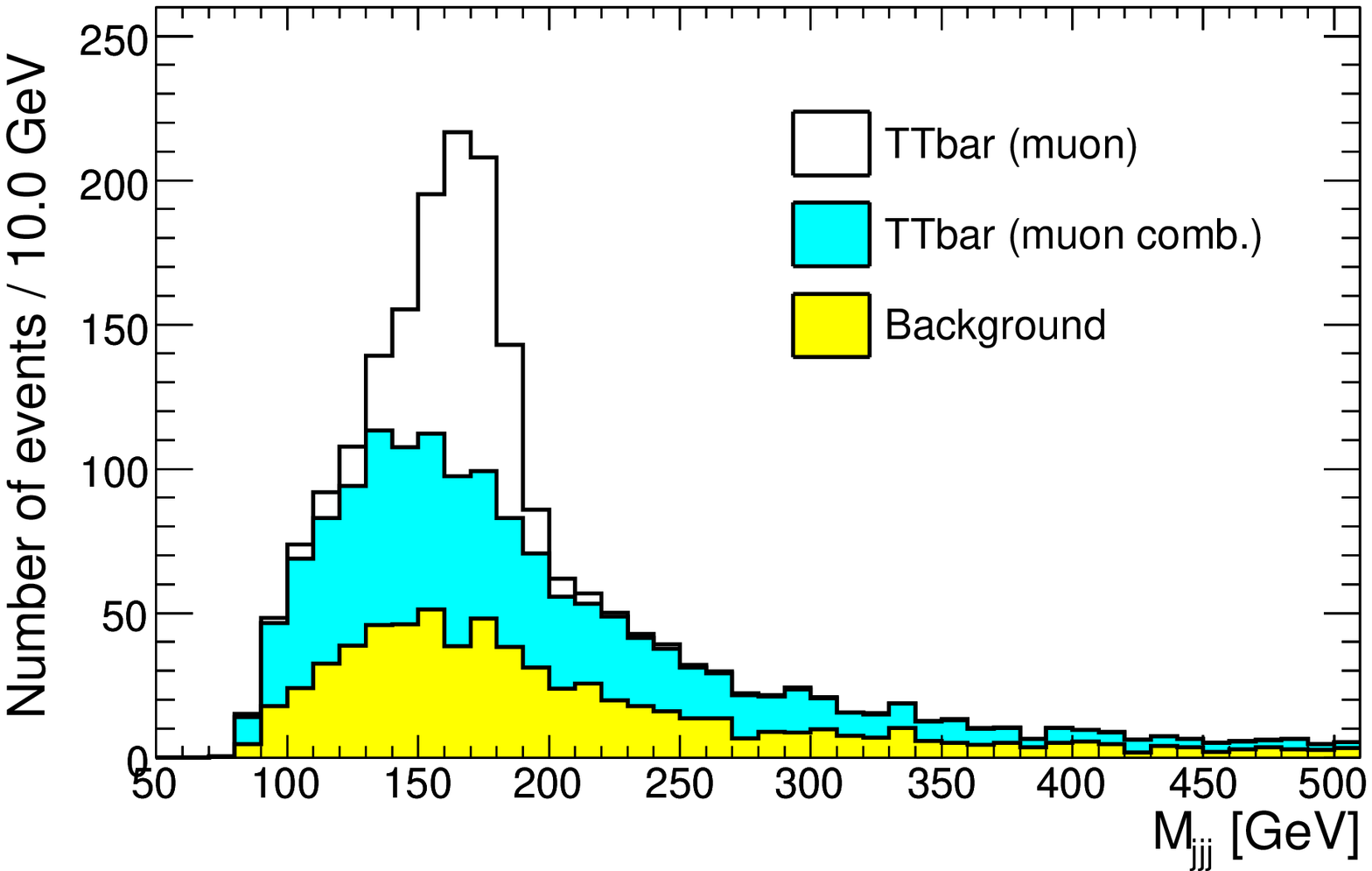}
\caption{Expectations for early top-pair observation. Left: CMS distribution 
of the number of jets after simple cuts in 10 pb$^{-1}$. Right ATLAS 
distribution of three-jet invariant mass after simple cuts in 100 pb$^{-1}$.}
\label{fig:toprediscovery}
\end{figure}

Early observation of $t \bar{t}$ production is possible both in the semi-leptonic
and in the di-leptonic channel. As an example in the semi-leptonic channel,
Fig.~\ref{fig:toprediscovery} (left) shows the expected distribution of the
number of jets after a few basic cuts in 10 pb$^{-1}$ of CMS data~\cite{cmstop}; 
the presence of a top signal is clear.

Figure~\ref{fig:toprediscovery} (right) shows for 100 pb$^{-1}$ of
ATLAS data the expected invariant mass of three hadronic jets (the three with
the highest $p_T$ vector sum) after simple cuts on one lepton, at least
four jets and some missing $E_T$, even without the use of $b$-tagging, which
may be compromised in early data. The main background is W+jets, but
the top signal is obvious. The cross-section can be extracted from
a fit to the invariant mass shape, or by background subtraction.
ATLAS expects in 100 pb$^{-1}$ a measurement of the $t \bar{t}$ 
cross-section with 3-7\% statistical uncertainty and 15\% systematic uncertainty,
not including the pdf error (3\%) and the luminosity error (5\%).

The di-leptonic channel has a smaller branching fraction, but less
background, and also here a signal in 10 pb$^{-1}$ seems feasible.

When well-understood $b$-tagging is available, the top signal becomes
very clean, and accurate cross-section measurements may be made.
Table~\ref{tab:cmssigmatop} shows the CMS estimates of uncertainties in
the $t \bar{t}$ cross-section in the semi-leptonic decay channel
(muons only). A major uncertainty is due to the error
on the $b$-tagging; 5\% is considered conservative and may well
be considerably smaller with 10 fb$^{-1}$ of luminosity.

\begin{table}[tbh]
\begin{center}
\caption{CMS estimates of uncertainties in the measurement of the
$t \bar{t}$ cross-section in the semi-leptonic decay channel
(muons only) for 1, 5 and 10 fb$^{-1}$.}
\label{tab:cmssigmatop}
\begin{tabular}{|l|c|c|c|} \hline
     & \multicolumn{3}{|c|}{$\Delta \sigma / \sigma$} \\
     & 1 fb$^{-1}$ & 5 fb$^{-1}$ & 10 fb$^{-1}$ \\ \hline
Simulation samples & \multicolumn{3}{|c|}{0.6\%} \\
Pile-up (30\% on-off) & \multicolumn{3}{|c|}{3.2\%} \\
Underlying event & \multicolumn{3}{|c|}{0.8\%} \\
Jet energy scale (light $q$, 2\%) & \multicolumn{3}{|c|}{1.6\%} \\
Jet energy scale (heavy $q$, 2\%) & \multicolumn{3}{|c|}{1.6\%} \\
Radiation ($\Lambda_{\mathrm{QCD}},Q_0^2$) & \multicolumn{3}{|c|}{2.6\%} \\
Fragmentation (Lund b, $\sigma_q$) & \multicolumn{3}{|c|}{1.0\%} \\
$b$-tagging (5\%) & \multicolumn{3}{|c|}{7.0\%} \\
Parton density functions & \multicolumn{3}{|c|}{3.4\%} \\
Background level & \multicolumn{3}{|c|}{0.9\%} \\
Integrated luminosity & 10\% & 5\% & 3\% \\ \hline
Statistical uncertainty & 1.2\% & 0.6\% & 0.4\% \\ \hline
Systematic uncertainty & 13.6\% & 10.5\% & 9.7\% \\ \hline
Total uncertainty & 13.7\% & 10.5\% & 9.7\% \\ \hline
\end{tabular}
\end{center}
\end{table}

\subsection{Single top production}

As mentioned, single top production takes place via three mechanisms,
each having its own dedicated analysis. The major backgrounds for all
three are top-quark pair production, multi-jet QCD and W+jets events.
In particular the QCD background can be suppressed by only looking at
leptonic decays of the W from the top.

\subsubsection{t-channel}

CMS performs a cut-based analysis of this channel, where cuts are
optimized with a genetic algorithm. In 10 fb$^{-1}$, the statistical
significance of the t-channel signal is 37, and the cross-section
is measured with a 2.7\% statistical error and 8\% systematic error
(excluding the luminosity uncertainty).

\begin{figure}[tbh]
\centering
\includegraphics[width=38mm,height=35mm]{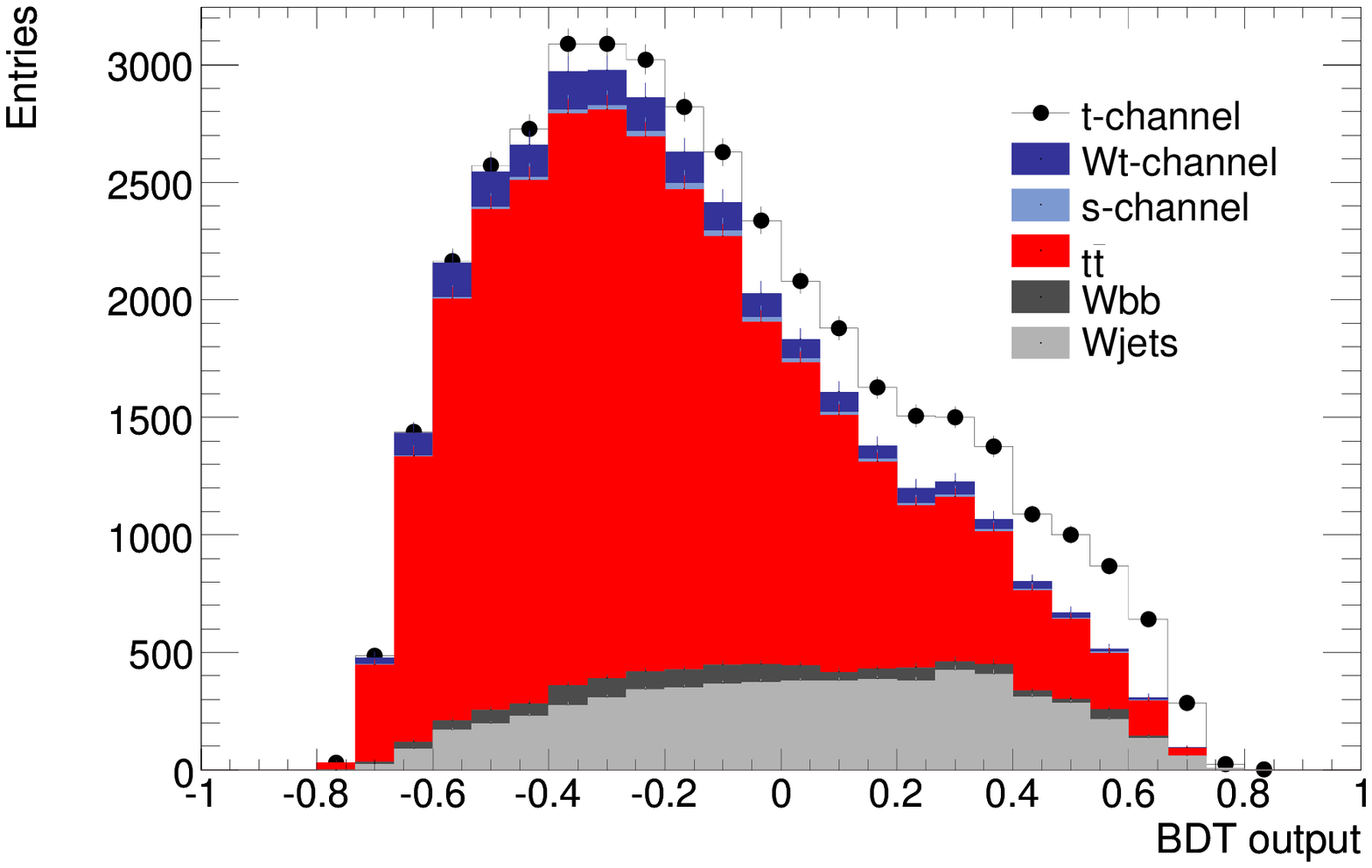}
\includegraphics[width=38mm,height=35mm]{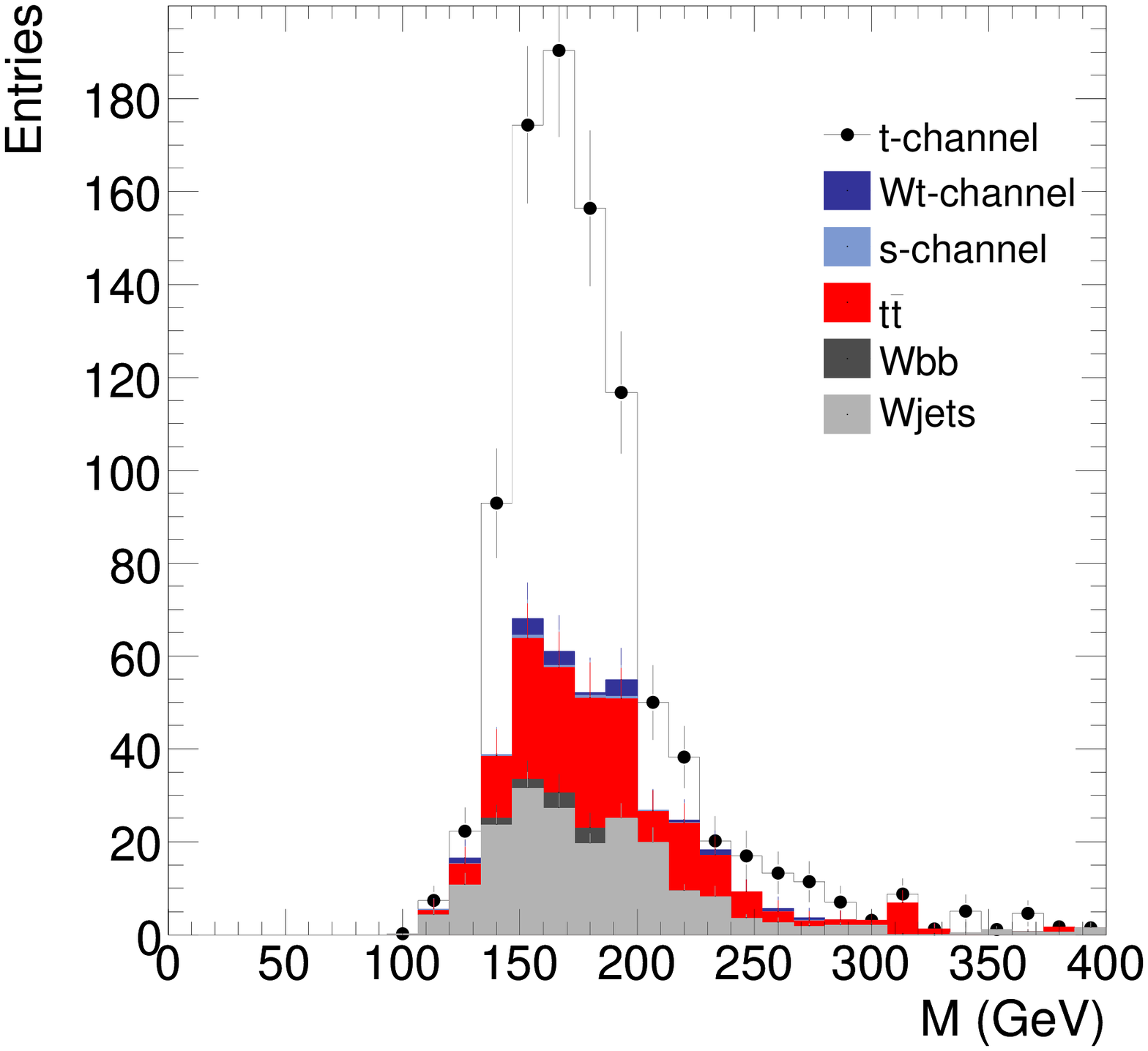}
\caption{Expectations for $t$-channel single top analysis. Left: ATLAS 
boosted decision tree (BDT) output. Right: reconstructed invariant mass of 
top decay products for BDT output larger than 0.6.}
\label{fig:singletopbdt}
\end{figure}

ATLAS uses a cut-based analysis as a baseline for robust observation,
but observes that a multivariate analysis using a boosted decision tree
(BDT) has a higher sensitivity. 
Figure~\ref{fig:singletopbdt} (left) shows the output of the BDT,
Fig.~\ref{fig:singletopbdt} (right) shows the invariant mass of the
top decay products after demanding that the BDT output is larger
than 0.6: the top peak is clear.

In 1 fb$^{-1}$, ATLAS aims to measure the
cross-section in the t-channel with 5.7\% statistical and 22\% systematic
error, which would determine $|V_{tb}|$ to $\pm 11$\% (stat.+syst.)
$\pm 4$\% (theory). In 10 fb$^{-1}$, the cross-section would be
measured to 2\% statistical and 10\% systematic errors.

\subsubsection{Associated tW production}

Here two final states are considered: the di-leptonic channel
$\ell^+ \ell^- E_T^{\mathrm{miss}} b$, and the semi-leptonic
channel $\ell^{\pm} E_T^{\mathrm{miss}} b j j$.

In 10 fb$^{-1}$ CMS expects a significance for a signal of $4.2$
in the di-leptonic channel, and $5.1$ in the semi-leptonic channel.
Excluding the systematic uncertainty due to limited MC statistics,
the cross-section in the di-leptonic channel is measured with a
statistical error of 8.8\% and a systematic error of 24\%; the 
cross-section in the semi-leptonic channel is measured with a
statistical error of 7.5\% and a systematic error of 17\%.
The dominating systematic errors are the jet energy scale, pile-up,
and $b$-tagging uncertainties.

ATLAS again uses cut-based and multivariate analyses, with in 10
fb$^{-1}$ an expected uncertainty on the measured cross-section
of 6.6\% statistical, and 19.4\% systematic.

\subsubsection{s-channel}

The s-channel is interesting since other particles like H$^{\pm}$ can
appear in the propagator, but it is a difficult channel due to the
low cross-section and large backgrounds.

CMS has studied this channel with a fast detector simulation, and
expects to reach a measurement of the cross-section in 10 fb$^{-1}$
with a 18\% statistical error, and a 31\% systematic error.

ATLAS has studied likelihood methods with full detector simulation,
and expects in 10 fb$^{-1}$ a statistical error of 20\% and a
systematic error of 48\% on the cross-section.

\section{Top properties}

It is expected that the LHC experiments can
make studies of top quark properties to levels exceeding those of
the Tevatron. However, for many measurements, in particular the top
mass, systematic errors dominate over statistical errors, and the full
capabilities of the LHC can only be exploited when those systematic
uncertainties are under control.

\subsection{Top mass}

The mass of the top quark is one of the most important parameters of
the Standard Model. It should be realized that, theoretically, the
concept of a pole mass for the top quark has an intrinsic uncertainty
of order $\Lambda_{\mathrm{QCD}}$, and a direct translation of the
parameter measured by experiments to the pole mass is not so obvious.
Perhaps the fairest statement one can make is that what experiments
measure is actually the ``top mass'' parameter in the Monte Carlo
generators...

CMS measures the top mass in the semi-leptonic channel using a full
kinematic fit to the events, and using the result of this fit in an 
event-by-event likelihood as a function of the top mass. This likelihood
can make optimal use of all information available in the event.
In a data sample of 10 fb$^{-1}$, CMS expects a statistical uncertainty
on the top mass of only 200 MeV, and a systematic uncertainty of 1.1 GeV if
the dominating uncertainty, the $b$-jet energy scale, is known to 1.5\%.

\begin{figure*}[tb]
\centering
\includegraphics[width=65mm]{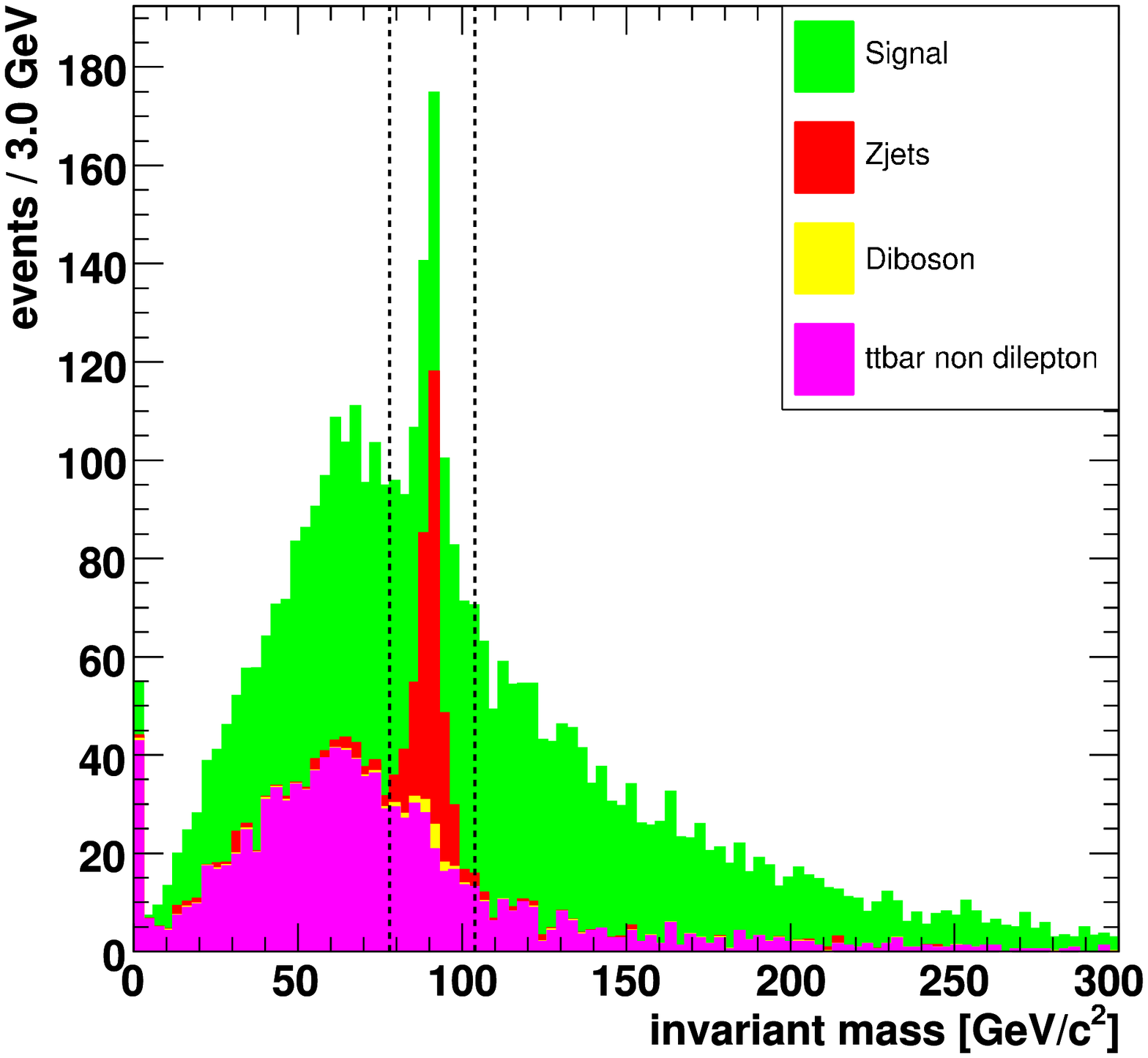}
\includegraphics[width=65mm]{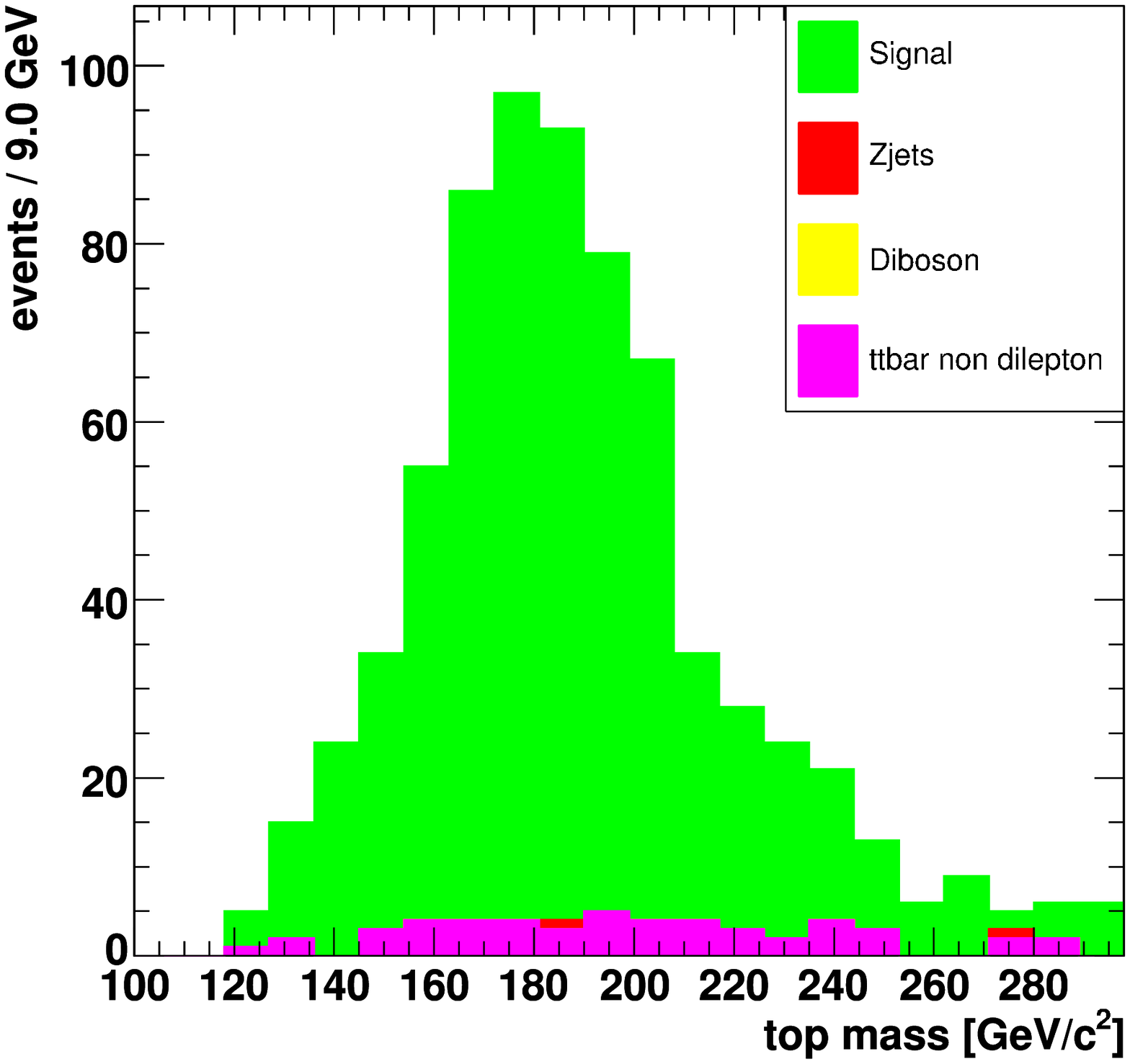}
\caption{Left: invariant mass of the two lepton candidates in di-leptonic
$t \bar{t}$ events in CMS, indicating the cut window to remove
Z+jets events. Right: most likely top mass after selection for 1 fb$^{-1}$.}
\label{fig:cmsllmtop}
\end{figure*}

CMS has also studied the extraction of the top mass in the di-lepton
channel and the all-hadronic channel. The di-lepton channel is very
clean, as shown in Fig.~\ref{fig:cmsllmtop}. Already in 1 fb$^{-1}$ a
measurement with an error of $\Delta m_t = 4.2$ GeV can be done,
improving to $\Delta m_t = 0.5$ GeV (stat.) $\pm 1.1$ GeV (syst.) in
10 fb$^{-1}$.
Also the all-hadronic channel contributes to the combined top mass;
its dominating systematics are the jet energy scale, radiation of
extra jets and the backgrounds.

The ATLAS studies have concentrated on the semi-leptonic decay channel,
with electrons and muons. Two algorithms have been studied to choose the
two light jets from the hadronically decaying W boson: a $\chi^2$
minimization method with event-by-event rescaling, and a geometric
method choosing the two light jets with smallest 
$\Delta R (= \sqrt{\Delta \phi^2 + \Delta \eta^2})$. 
Then the two $b$-tagged jets are assigned to hadronic and leptonic W to
form the two top quarks. The top quark mass can be extracted from
a fit to the invariant mass peak, or from a full kinematic fit of the
event. In the final expected result for 1 fb$^{-1}$, the systematic 
uncertainty dominates over the statistical one, and amounts to
0.7 GeV per \% of $b$-jet energy scale uncertainty, 0.2 GeV per \%
of light jet energy scale uncertainty, and $\sim 0.3$ GeV due to
uncertainties related to radiation of extra jets, either
from the initial (ISR) or final state (FSR).

An interesting alternative method to measure the top mass has been
studied by CMS, and involves
selection of events where a $b$-quark decays into a $J/\Psi$ ($+X$) and
the $J/\Psi$ into two leptons, and where the W from the same top quark also decays
leptonically. The invariant mass of the three leptons is sensitive to the top 
mass; the systematic uncertainties of this method include $b$-decay modelling
and the lepton energy scale, but not the $b$-jet energy scale, and it is thus
almost orthogonal to the standard methods. In 20 fb$^{-1}$ the statistical
error could reach $\sim 1$ GeV and the systematic error $\sim 1.5$ GeV,
dominated by theory systematics that may be further reduced by new
calculations.

\subsection{Top decay}

If the top quark is indeed the $Q = 2/3$, spin $1/2$, heavy partner of
the bottom quark, then the following expression holds for its decay width to 
a W and a lighter quark $q$ ($q = d,s,b$):
\begin{eqnarray}
\Gamma (t \to W^+ q) & = & \frac{G_F |V_{tq}|^2 m_t^3}{8 \pi \sqrt{2}}
(1 - \frac{m_W^2}{m_t^2})^2 
 (1 + 2 \frac{m_W^2}{m_t^2})  \nonumber \\
 & \times & (1 - 0.81 \alpha_s - 1.8 \alpha_s^2).
\end{eqnarray}

In the Standard Model $|V_{tb}| \approx 1$, and the top decay into $b$W 
dominates far above $s$W and $d$W decays. The decay has the structure of the
typical charged weak V-A (vector minus axial vector coupling) form. 
It is possible to measure $|V_{tb}|$ by counting the number of $b$-tagged 
jets in top decays; the precision will be determined by $b$-tagging systematics.
Alternatively, $|V_{tb}|$ can also be determined in single top production.

\subsubsection{Top charge}

For a decay $t \to W b$, it is interesting to study the correlation of
W and $b$ charge, in order to establish whether indeed $t \to W^+ b$ is
observed as expected in the SM, or whether we are observing an exotic
``top-like'' quark with $Q = 4/3$ decaying as $t \to W^+ \bar{b}$.
At the Tevatron, this latter scenario is disfavoured at the 90-95\% CL.
ATLAS can make a $5 \sigma$ distinction with 1 fb$^{-1}$ with a $b$-jet
charge technique, or using semi-leptonic $b$-decays.

\subsubsection{FCNC decays} 

In the SM, flavour-changing neutral current top decays ($t \to q$ X, where 
$q = c,u$ and X $ = \gamma$,Z,gluon) are strongly suppressed
(${\cal{O}}(10^{-14})$); in certain models of new physics they can be 
enhanced to levels of $10^{-4} - 10^{-5}$. 

\begin{figure}[tbh]
\centering
\includegraphics[width=70mm]{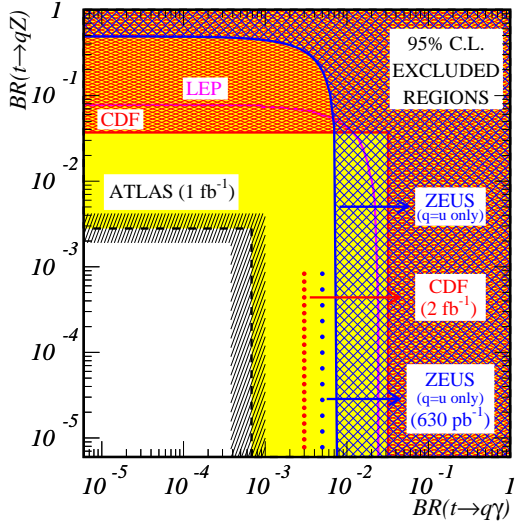}
\caption{ATLAS expectation for 95\% CL limits on the branching ratios
of the FCNC decays $t \to q \gamma$ and $t \to q Z$ in 1 fb$^{-1}$.
\label{fig:atlasfcnc}}
\end{figure}

CMS has evaluated the sensitivity to X $ = \gamma$, Z with a cut-based
analysis. In 10 fb$^{-1}$, a $5 \sigma$ discovery of Br$(t \to q \gamma)
> 8.4 \times 10^{-4}$ and Br$(t \to q Z) > 15 \times 10^{-4}$ could
be made. ATLAS has developed a multivariate selection for X $= \gamma$, Z 
and gluon, and quotes the results as 95\% CL exclusion limits for 
1 fb$^{-1}$, as shown in Fig.~\ref{fig:atlasfcnc}.

\subsubsection{V-A structure}

The V-A structure of top quark decay can be studied in semi-leptonic
$t \bar{t}$ decays. One measures the polarisation of the leptonically
decaying W from the distribution of the angle $\Psi$ between the lepton in 
the W frame, and the W in the top frame. This distribution can be written as
\begin{eqnarray}
\frac{1}{N} \frac{dN}{d \cos \Psi} & = & \frac{3}{2} [ F_0 
(\frac{\sin \Psi}{\sqrt{2}} )^2  \nonumber \\
 & \! \! \! + & \! \! \! F_L ( \frac{1 - \cos \Psi}{2})^2
+ F_R ( \frac{1 + \cos \Psi}{2} )^2 ] ,
\end{eqnarray}
where $F_0$, $F_L$ and $F_R$ are the fractions of longitudinally, left-handed
and right-handed polarized W's, respectively. By fitting the appropriate
$\Psi$-dependent functions to the distribution, the fractions can be
extracted, under the constraint $F_0 + F_L + F_R = 1$. 

In 730 pb$^{-1}$, ATLAS expects to measure, if indeed these fractions
are as the Standard Model predicts: $F_0 = 0.70 \pm 0.04 \pm 0.02$,
$F_L = 0.29 \pm 0.02 \pm 0.03$, and $F_R = 0.01 \pm 0.02 \pm 0.02$, where
the first error is statistical and the second systematic.

\subsubsection{Anomalous couplings}

In a more general way, it is convenient to write an effective Lagrangian
for the $tbW$ vertex as follows: 
\begin{eqnarray}
{\cal{L}} & = & - \frac{g}{\sqrt{2}} \bar{b} \gamma^{\mu}
(V_L P_L + V_R P_R) t W_{\mu}^-  \nonumber \\
 & - & \frac{g}{\sqrt{2}} \bar{b}
\frac{i \sigma^{\mu \nu} q_{\nu}}{M_W} (g_L P_L + g_R P_R)
t W_{\mu}^-  \, (\mathrm{+ h.c.}) ,
\end{eqnarray}
where $P_{R/L} = (1 \pm \gamma^5)/2$ are the usual right- and left-handed
projection operators, and $V_{R/L}$ and $g_{R/L}$ are top couplings; in
the SM only $V_L \neq 0$ (in fact $V_L = V_{tb}$), and the other couplings
are anomalous.

\begin{figure*}[tb]
\centering
\includegraphics[width=65mm]{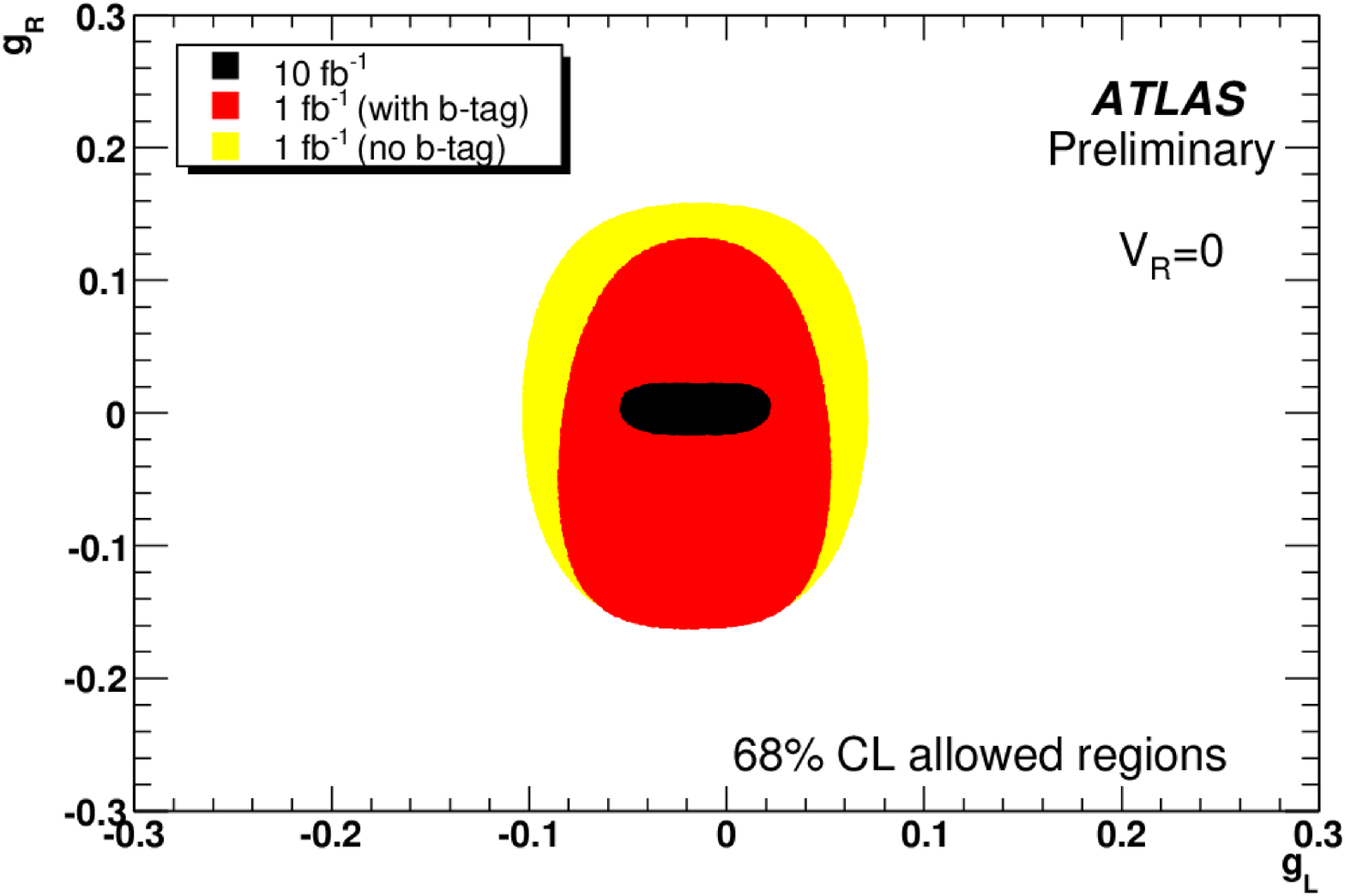}
\includegraphics[width=65mm]{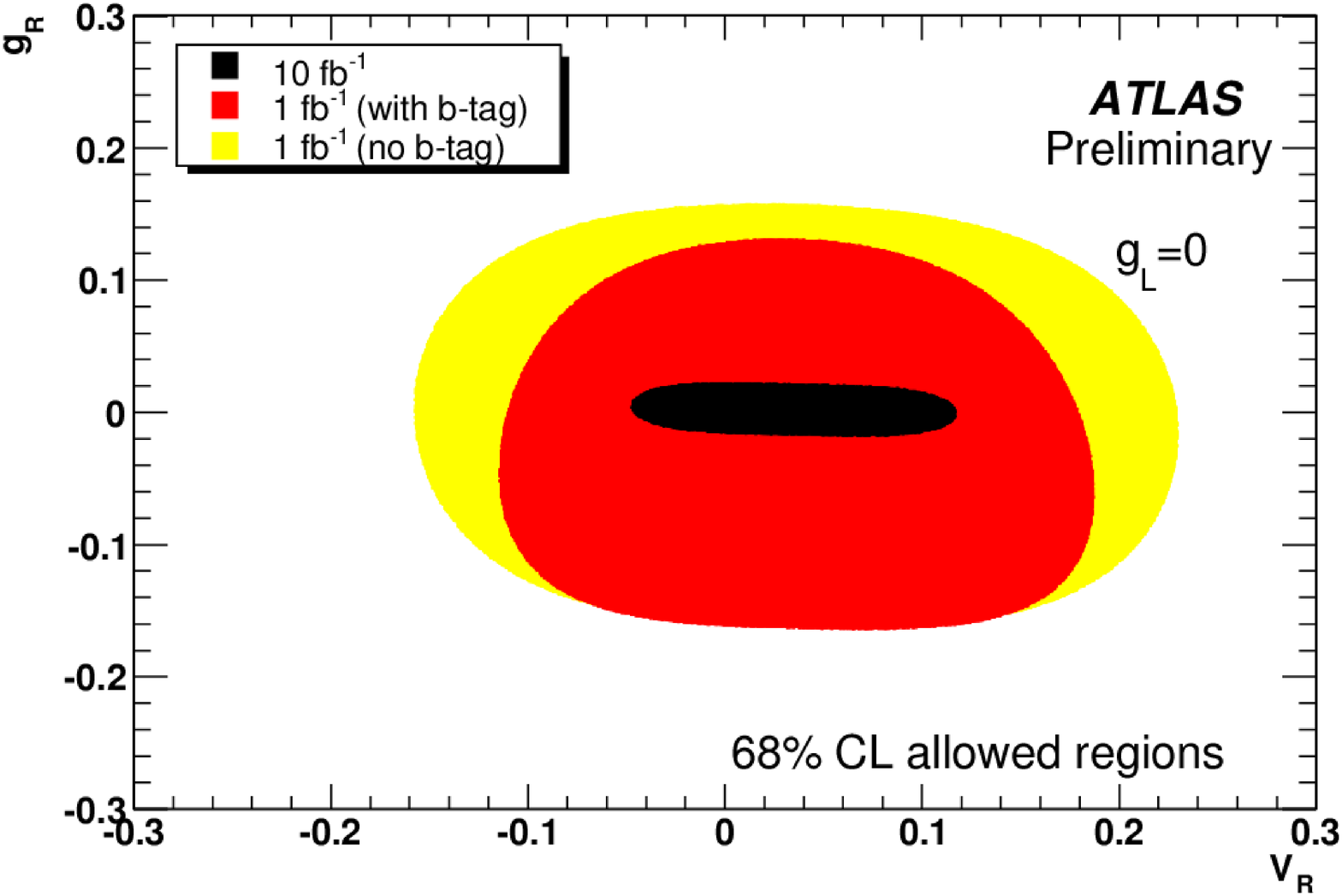}
\caption{Expected limits on the anomalous tbW couplings $g_L$ versus $g_R$
(left) and $V_R$ versus $g_R$ (right), in 1 and 10 fb$^{-1}$ with and
without $b$-tagging, derived by ATLAS.} 
\label{fig:atlasac}
\end{figure*}

ATLAS has derived expected limits on the anomalous couplings for
1 and 10 fb$^{-1}$, as shown in Fig.~\ref{fig:atlasac}.

\subsection{Spin correlations}

\begin{figure}[tbh]
\centering
\includegraphics[width=38mm]{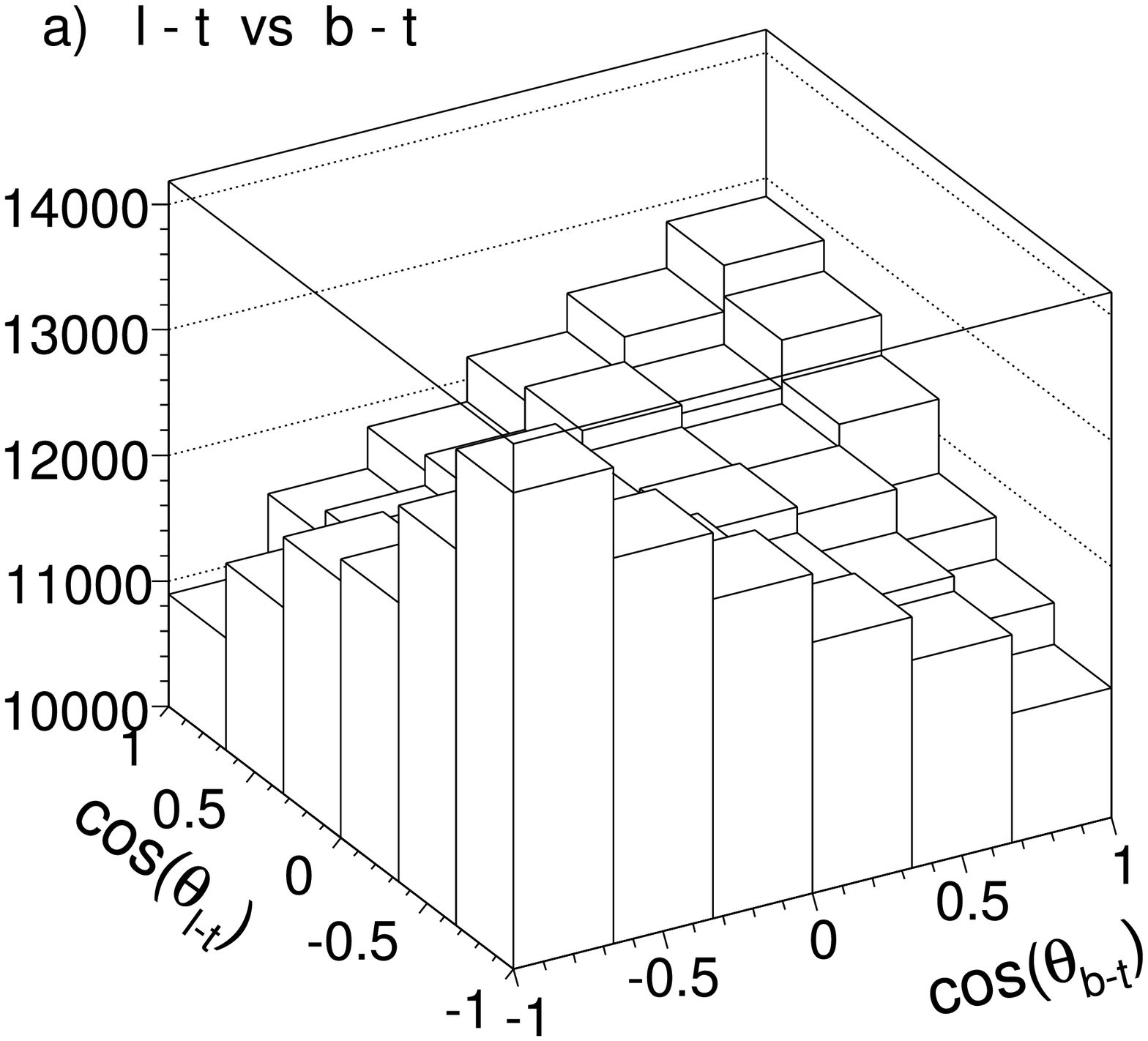}
\includegraphics[width=38mm]{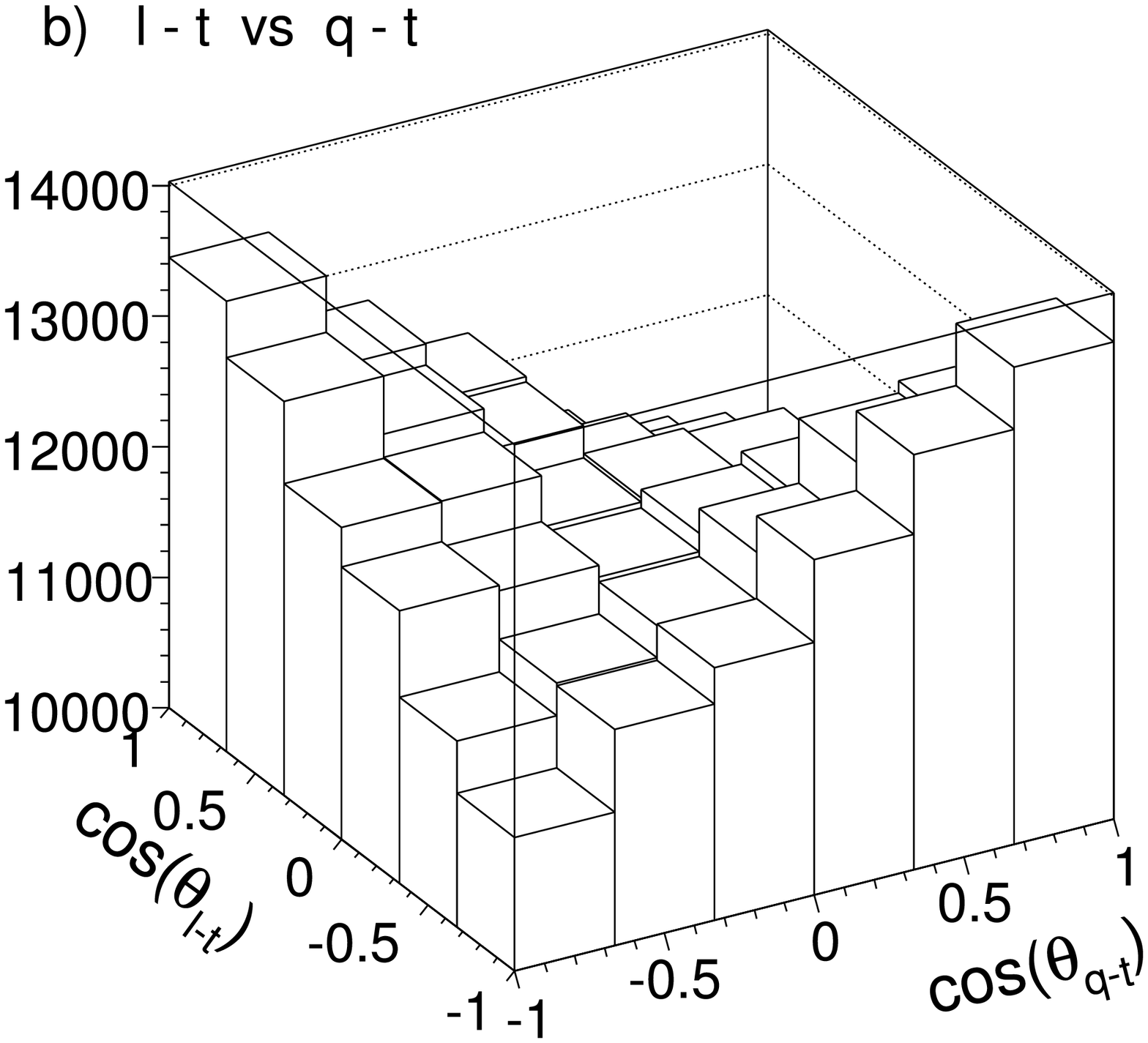}
\caption{Distributions of of $\cos \theta_1$ versus $\cos \theta_2$ for
$\theta_1 = \theta_{l-t}, \theta_2 = \theta_{b-t}$ (left) and
$\theta_1 = \theta_{l-t}, \theta_2 = \theta_{q-t}$ (right), for
CMS before detector simulation.}
\label{fig:spincor}
\end{figure}

The top itself is expected to be produced essentially unpolarised, but
there are correlations between the spins of the two tops in the same
event. This is due to the fact that close to threshold, in gluon-gluon
fusion the top-pairs are produced in a $^1 S_0$ state. (In $q 
{\bar{q}}^{\prime}$
annihilation they would be in a $^3 S_1 $ state). The top spin correlations 
can be studied by looking at the asymmetry $A$ between parallel aligned top spins 
versus oppositely aligned top spins
\begin{equation}
A = \frac{\sigma(\uparrow \uparrow) + \sigma(\downarrow \downarrow) -
\sigma(\uparrow \downarrow) - \sigma(\downarrow \uparrow)}
{\sigma(\uparrow \uparrow) + \sigma(\downarrow \downarrow) +
\sigma(\uparrow \downarrow) + \sigma(\downarrow \uparrow)},
\end{equation}
where $\sigma(\uparrow \downarrow)$ stands for the cross-section of
producing a $t$ with spin up and a $\bar{t}$ with spin down, etc.
As a spin-analyzer axis we use the direction of flight of the two tops 
in their combined center-of-mass frame. Defining $\theta_i$ as the
angle between $t$ (or $\bar{t}$) decay product $i$ in the $t$ (or
$\bar{t}$) rest frame, and the $t$ ($\bar{t}$) in the $t \bar{t}$ rest frame,
the double differential distribution of $\theta_1$ and $\theta_2$,
where $1$ and $2$ are from different top quarks, can be written as:
\begin{eqnarray}
\frac{1}{N} \frac{d^2 N}{d \cos \theta_1 d \cos \theta_2} = &  \nonumber \\
 & \! \! \! \! \! \! \frac{1}{4}(1 - A |\alpha_1 \alpha_2| \cos \theta_1
\cos \theta_2).
\end{eqnarray}
In this equation, $\alpha_i$ is the spin analyzing power of decay
product $i$: it is nearly one for a lepton or a $d$-type quark if one
could identify such a quark, it is typically $0.5$ for tagged $b$-quark
jets or the lowest energy jet. Figure~\ref{fig:spincor} shows the
distributions of $\cos \theta_1$ versus $\cos \theta_2$ for
$\theta_1 = \theta_{l-t}, \theta_2 = \theta_{b-t}$ (left) and
$\theta_1 = \theta_{l-t}, \theta_2 = \theta_{q-t}$ (right), for
CMS at generator level, before detector simulation. Detector effects do
significantly deteriorate the correlations and need to be taken into
account, leading to systematic uncertainties. For top-quark production dominated by 
gluon-gluon fusion, $A$ is expected to be $0.3 - 0.4$ depending on exact 
cuts, whereas for $t \bar{t}$ production by quark-antiquark annihilation $A<0$.

CMS expects to measure $A$ in 10 fb$^{-1}$ with a total relative uncertainty
(dominated by systematics) of 20-25\%, which would clearly distinguish
between production mechanisms. ATLAS expects a measurement with a relative
uncertainty of ${\cal{O}}(50)$\% already in 200 pb$^{-1}$, evolving to
10-20\% in 10 fb$^{-1}$.

\section{Anomalous production}

New physics can introduce new production mechanisms for single top quarks or 
top quark pairs. Flavour-changing neutral current processes can lead to
anomalous single top production. Furthermore, top quarks can be produced
in the decay of stop quarks in supersymmetry, in charged Higgs decays,
or in the decay of resonances that appear in little Higgs or LR-symmetric Higgs
models, or in models with extra dimensions containing Kaluza-Klein (KK) 
states. Apart from a deviation in the overall single-top or top-pair
cross-section, a pronounced deviation in the top-pair invariant mass
distribution may appear. A general complication in the reconstruction of
top quarks produced in the decay of massive particles is the fact that
these top quarks are highly boosted, and their decay products are very
close. Dedicated reconstruction techniques are under development in
both CMS and ATLAS.

\section{Conclusion}

The LHC will provide a very large sample of top quarks. These top
quarks play a central role in many studies, and statistical errors
will not be the limiting factor (with the exception of rare decays).
Detailed studies of systematic errors, on the other hand,
will be very important. On the detector side these include
$b$-jet and light jet energy scales, $b$-tagging, and pile-up. On the theory
side, the major uncertainties are from radiation of extra jets 
(ISR/FSR), pdf's, and hadronization.
Hard work will be needed by experimentalists and theorists alike, with
the reward of better understanding of a very special quark.

\bigskip 
\begin{acknowledgments}
The author wishes to thank everyone in CMS and ATLAS whose work
contributed to this talk.
\end{acknowledgments}

\bigskip 

\end{document}